\documentclass[preprint,amsmath,amssymb,aps,nofootinbib,showpacs]{revtex4}
\usepackage{graphicx}
\usepackage{bm}
\usepackage{subfigure}
\usepackage{amsmath}

\newcommand{\epl}{Europhys. Lett.\ }

\newcommand{\pr}{Phys. Rev.\ }

\newcommand{\jpa}{J. Phys. A\ }
\newcommand{\jpb}{J. Phys. B\ }

\newcommand{\etal}{{\em et al. }}

\newcommand{\UQ}{School of Mathematics and Physics, University of Queensland, Brisbane, 
QLD 4072, Australia.}

\begin{document}

\title{Spreading of entanglement and steering along small Bose-Hubbard chains}

\author{M.~K. Olsen}
\affiliation{\UQ}
\date{\today}

\begin{abstract}

We investigate how entanglement spreads along small Bose-Hubbard chains, with only the first well initially occupied by a mesoscopic number of atoms, as the number of sites increases. For two- and three-well chains in the non-interacting case, we are able to obtain analytical solutions and show that the presence of entanglement depends on having a sub-Poissonian state of the atoms in the first well. In these cases, the correlations we calculate are completely periodic. Restoring the collisional interactions or moving to a four-well chain necessitates a numerical treatment, for which we use the fully quantum positive-P representation. We examine two different correlations and find that adding collisional interactions destroys the periodicity of the correlations and causes them to degrade with time. This happens well before there is a noticeable effect on the periodicity of the solutions for the number of atoms in each well.

\end{abstract}


\maketitle

\section{Introduction}
\label{sec:intro}

The investigation of non-local correlations such as entanglement and Einstein-Podolsky-Rosen (EPR) steering~\cite{Einstein,Erwin,Wiseman} necessarily belongs to the
field of continuous-variable entanglement, which is a very active area of research~\cite{Braunstein,Stefano}.
Many criteria having been developed to signify the presence of inseparability, entanglement and the potential for EPR steering, especially in bipartite systems. 
The measures that we use in this work were developed from work by Hillery and Zubairy~\cite{HZ} which was based on the Cauchy-Schwarz inequality. The original inequality covered bipartite inseparability and was expanded on by Cavalcanti \etal\cite{ericsteer} to cover multipartite entanglement, steering, and violations of Bell inequalities. As shown by He \etal\cite{He}, the Hillery and Zubairy criteria are well suited to number conserving processes such as those of interest here, but as shown by Olsen~\cite{clervie2}, they will sometimes miss actually existing entanglement in quantum optical systems. In previous work on a different three-well system, we have found these to be the most useful criteria for the detection of entanglement in Bose-Hubbard systems~\cite{toberejected,BS3expand} 

Multi-mode entanglement in Bose-Einstein condensates (BEC) has been predicted and examined in the processes of molecular dissociation~\cite{KVK}, four-wave mixing in an optical lattice~\cite{Campbell,Mavis,Andy4}, and in a two-well Bose-Hubbard model~\cite{He,Hines}. In the latter case the separation of the modes is produced by the tunneling between wells, in both the 
continuous~\cite{Oberthaler2008,Oberthaler2011,He} and pulsed tunneling configurations~\cite{myJPB,myJOSAB}. The spreading of correlations and entanglement after a sudden parameter change, investigated by Lauchli and Kollath~\cite{quench}, and the development of entanglement between the two ends of a chain, analysed by Reslen and Bose~\cite{Reslen}, are particularly relevant to the present work.

 The quantum correlations necessary to detect inseparability, entanglement and EPR steering can in principle be measured using the interaction with light~\cite{HomoSimon}, or by homodyning with other atomic modes~\cite{HomoAndy}. 
We note here that the entanglement we are examining is a collective property between atomic modes which are spatially separated, and is not between individual atoms~\cite{Hines,Mavis}. We also note that tunneling of atoms between wells will not, by itself, lead to entanglement, with either a nonlinear interaction or initial non-classical states being necessary~\cite{superobvious}.

What we investigate in this article is the spreading of entanglement and the potential for EPR steering along small Bose-Hubbard chains, beginning with a mesoscopic population of atoms in one end well. In the case of non-interacting atoms and two- or three-well chains, we will obtain analytical solutions of the Heisenberg equations of motion for the atomic annihilation and creation operators. In these cases, we find that the presence of entanglement or EPR steering will depend on the presence of a non-classical quantum state of the initial atoms. For four wells, and in all cases with atomic interactions, we will use numerical stochastic integration in the positive-P representation~\cite{Pplus}, which proved to be stable over the evolution times we examine here. 

\section{Physical model and Hamiltonians}
\label{sec:model}

We follow the approach to the Bose-Hubbard model~\cite{BHmodel,Jaksch} taken by Milburn \etal~\cite{BHJoel}, also generalising this to three~\cite{Nemoto,Chiancathermal} and four wells~\cite{Chianca4}. All our systems are in a linear configuration with only the first well initially occupied. Where possible, we analytically solve the Heisenberg equations of motion and in other cases we use the fully quantum positive-P representation~\cite{Pplus}, which contains the full dynamics due to the Hamiltonians. We consider these to be the most suitable approaches here because they are both exact, allow for an easy representation of mesoscopic numbers of atoms, can be used to calculate quantum correlations, and can simulate different quantum initial states~\cite{states}. Just as importantly, both calculations scale linearly with the number of sites and can in principle deal with any number of atoms. One disadvantage of the positive-P representation is that the integration can show a tendency to diverge at short times for high collisional nonlinearities~\cite{Steel}. As long as the procedures followed to derive the Fokker-Planck equation for the positive-P function are valid~\cite{SMCrispin}, the stochastic solutions are guaranteed to be accurate wherever the integration converges. With all the results shown here, we found no sign of numerical divergences.

The systems we investigate are very simple, with the potential wells in a linear configuration. Each of these can contain a single atomic mode, which we will treat as being in the lowest energy level. Atoms in each of the wells can tunnel into the nearest neighbour potential.
With the $\hat{a}_{j}$ as bosonic annihilation operators for atoms in mode $j$, $J$ representing the coupling between the wells, and $\chi$ as the collisional nonlinearity, we may now write our Hamiltonians.
Following the usual procedures~\cite{BHJoel}, we find the two-well Hamiltonian as
\begin{equation}
{\cal H}_{2} = \hbar\sum_{j=1}^{2}\chi\hat{a}_{j}^{\dag\;2}\hat{a}_{j} + \hbar J\left(\hat{a}_{1}\hat{a}_{2}^{\dag}+\hat{a}_{1}^{\dag}\hat{a}_{2}  \right),
\label{eq:Ham2}
\end{equation}
the three-well Hamiltonian as
\begin{equation}
{\cal H}_{3} =  \hbar\sum_{j=1}^{3}\chi \hat{a}_{j}^{\dag\;2}\hat{a}_{j}^{2}
+\hbar J\left(\hat{a}_{1}^{\dag}\hat{a}_{2}+\hat{a}_{2}^{\dag}\hat{a}_{1}+\hat{a}_{3}^{\dag}\hat{a}_{2}
+\hat{a}_{2}^{\dag}\hat{a}_{3}\right),
\label{eq:Ham3}
\end{equation}
and that for the four well system as
\begin{equation}
{\cal H}_{4} =  \hbar\sum_{j=1}^{4}\chi \hat{a}_{j}^{\dag\;2}\hat{a}_{j}^{2}
+\hbar J\left(\hat{a}_{1}^{\dag}\hat{a}_{2}+\hat{a}_{2}^{\dag}\hat{a}_{1}+\hat{a}_{3}^{\dag}\hat{a}_{2}
+\hat{a}_{2}^{\dag}\hat{a}_{3}+\hat{a}_{4}^{\dag}\hat{a}_{3}+\hat{a}_{3}^{\dag}\hat{a}_{4}\right).
\label{eq:Ham4}
\end{equation}
From these three Hamiltonians we can derive the equations of motion used to investigate the dynamics of our systems. 

\section{Non-interacting case}
\label{sec:ernest}

For the case where the collisional interaction between the atoms is set to zero, we find that an analytical solution of the Heisenberg equations of motion of the system operators for the two- and three-well chains is possible. For the four-well chain, we were unsuccessful. The two-well system is rather simple and has been extensively analysed. We will present the results here for completeness and for the purposes of easy comparison with the longer chains.

\subsection{Two wells}
\label{subsec:twin}

For the two-well system, the Heisenberg equations of motion are found as 
\begin{equation}
\frac{d}{dt}\left[ \begin{array}{c}
\hat{a}_{1} \\ \hat{a}_{1}^{\dag} \\ \hat{a}_{2} \\ \hat{a}_{2}^{\dag} \end{array}\right]
=
\begin{bmatrix} 
0 & 0 & -iJ & 0 \\ 
0 & 0 & 0 & iJ  \\
-iJ & 0 & 0 & 0  \\
0 & iJ & 0 & 0  \\
\end{bmatrix}
\times \left[ \begin{array}{c}
\hat{a}_{1}(0) \\ \hat{a}_{1}^{\dag}(0) \\ \hat{a}_{2}(0) \\ \hat{a}_{2}^{\dag}(0)
\end{array} \right].
\label{eq:Heisenberg2}
\end{equation} 
This set of linear operator equations is readily solved, having the solutions
\begin{eqnarray}
\hat{a}_{1}(t) &=& \hat{a}_{1}(0)\cos Jt - i\hat{a}_{2}(0)\sin Jt, \nonumber \\
\hat{a}_{1}^{\dag}(t) &=&\hat{a}_{1}^{\dag}(0)\cos Jt + i\hat{a}_{2}^{\dag}(0)\sin Jt, \nonumber \\
\hat{a}_{2}(t) &=& -i\hat{a}_{1}(0)\sin Jt+\hat{a}_{2}(0)\cos Jt , \nonumber \\
\hat{a}_{2}^{\dag}(t) &=&  i\hat{a}_{1}^{\dag}(0)\sin Jt+\hat{a}_{2}^{\dag}(0)\cos Jt
\label{eq:Hsols2}
\end{eqnarray}

For the populations, we find the time dependent solutions,
\begin{eqnarray}
\langle\hat{N}_{1}(t)\rangle &=& \langle\hat{a}_{1}^{\dag}(0)\hat{a}_{1}(0)\rangle \cos^{2}Jt,\nonumber\\
\langle\hat{N}_{2}(t)\rangle &=& \langle\hat{a}_{1}^{\dag}(0)\hat{a}_{1}(0)\rangle \sin^{2}Jt,
\label{eq:Ernienum2}
\end{eqnarray}
showing that the atoms will cycle forever between the two wells, as long as any other considerations are absent. The inclusion of the atomic interactions is known to cause a collapse and eventual revival of the oscillations~\cite{Chiancathermal}, but over longer timescales than we are investigating here.

\subsection{Triple wells}
\label{subsec:triple}

For the non-interacting three-well system, the Heisenberg equations of motion are
\begin{equation}
\frac{d}{dt}\left[ \begin{array}{c}
\hat{a}_{1} \\ \hat{a}_{1}^{\dag} \\ \hat{a}_{2} \\ \hat{a}_{2}^{\dag} \\ \hat{a}_{3} \\ \hat{a}_{3}^{\dag} \end{array}\right]
=
\begin{bmatrix} 0 & 0 & -iJ & 0 & 0 & 0 \\ 
0 & 0 & 0 & iJ & 0 & 0 \\
-iJ & 0 & 0 & 0 & -iJ & 0 \\
0 & iJ & 0 & 0 & 0 & iJ \\
0 & 0 & -iJ  & 0 & 0 & 0 \\
0 & 0 & 0 & iJ & 0 & 0
\end{bmatrix}
\times \left[ \begin{array}{c}
\hat{a}_{1}(0) \\ \hat{a}_{1}^{\dag}(0) \\ \hat{a}_{2}(0) \\ \hat{a}_{2}^{\dag}(0) \\ \hat{a}_{3}(0) \\ \hat{a}_{3}^{\dag}(0)
\end{array} \right].
\label{eq:Heisenberg3}
\end{equation} 

\begin{figure}
\begin{center}
\includegraphics[width=0.8\columnwidth]{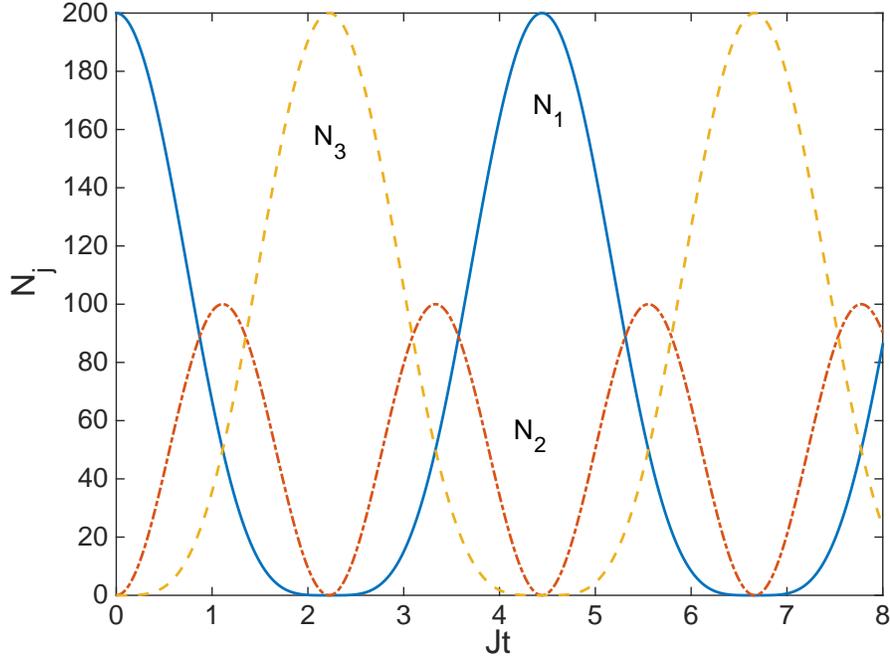}
\end{center}
\caption{(Color online) The analytical non-interacting solutions for the populations in each well as a function of time, for $J=1$ and $N_{1}(0)=200$, with $N_{2}(0)=N_{3}(0)=0$. The interacting stochastic results for $\chi=10^{-3}$ and both coherent and Fock initial states are indistinguishable on this scale. The quantities plotted in this and subsequent plots are dimensionless.}
\label{fig:populations3}
\end{figure}

This set of linear operator equations is also readily solved, having the solutions
\begin{eqnarray}
\hat{a}_{1}(t) &=& \frac{1}{2}\left(\cos \Omega t + 1 \right)\hat{a}_{1}(0) - \frac{i}{\sqrt{2}}\sin\Omega t\: \hat{a}_{2}(0) + \frac{1}{2}\left(\cos\Omega t - 1\right)\hat{a}_{3}(0), \nonumber \\
\hat{a}_{1}^{\dag}(t) &=& \frac{1}{2}\left(\cos \Omega t + 1 \right)\hat{a}_{1}^{\dag}(0) + \frac{i}{\sqrt{2}}\sin\Omega t\: \hat{a}_{2}^{\dag}(0) + \frac{1}{2}\left(\cos\Omega t - 1\right)\hat{a}_{3}^{\dag}(0), \nonumber \\
\hat{a}_{2}(t) &=& \frac{-i}{\sqrt{2}}\sin\Omega t\: \hat{a}_{1}(0) + \cos\Omega t\:\hat{a}_{2}(0) - \frac{i}{\sqrt{2}}\sin\Omega t\:\hat{a}_{3}(0), \nonumber \\
\hat{a}_{2}^{\dag}(t) &=& \frac{i}{\sqrt{2}}\sin\Omega t\: \hat{a}_{1}^{\dag}(0) + \cos\Omega t\:\hat{a}_{2}^{\dag}(0) + \frac{i}{\sqrt{2}}\sin\Omega t\:\hat{a}_{3}^{\dag}(0), \nonumber \\
\hat{a}_{3}(t) &=& \frac{1}{2}\left(\cos \Omega t - 1 \right)\hat{a}_{1}(0) - \frac{i}{\sqrt{2}}\sin\Omega t \:\hat{a}_{2}(0) + \frac{1}{2}\left(\cos\Omega t + 1\right)\hat{a}_{3}(0), \nonumber \\
\hat{a}_{3}^{\dag}(t) &=& \frac{1}{2}\left(\cos \Omega t - 1 \right)\hat{a}_{1}^{\dag}(0) + \frac{i}{\sqrt{2}}\sin\Omega t\: \hat{a}_{2}^{\dag}(0) + \frac{1}{2}\left(\cos\Omega t + 1\right)\hat{a}_{3}^{\dag}(0),
\label{eq:Hsols}
\end{eqnarray}
where we have made the substitution $\Omega = \sqrt{2} J$ for reasons of notational elegance. In this case we find the populations as
\begin{eqnarray}
\langle\hat{N}_{1}(t)\rangle &=& \frac{1}{4}\left(\cos\Omega t+1 \right)^{2}\langle\hat{a}_{1}^{\dag}(0)\hat{a}_{1}(0)\rangle,\nonumber \\
\langle\hat{N}_{2}(t)\rangle &=& \frac{1}{2}\sin^{2}\Omega t \langle\hat{a}_{1}^{\dag}(0)\hat{a}_{1}(0)\rangle, \nonumber \\
\langle\hat{N}_{3}(t)\rangle &=& \frac{1}{4}\left(\cos\Omega t-1 \right)^{2}\langle\hat{a}_{1}^{\dag}(0)\hat{a}_{1}(0)\rangle.
\label{eq:Ernienum3}
\end{eqnarray}
These solutions again describe fully periodic oscillations and are shown in Fig.~\ref{fig:populations3}.

\subsection{Four wells}
\label{subsec:BH4}

In the four well case, we were not able to obtain analytical solutions of the Heisenberg equations of motion. However, we can integrate the positive-P representation~\cite{Pplus} equations numerically with initial states taken from distributions using the methods developed by Olsen and Bradley~\cite{states}. For an initial coherent state, this is the same as integrating the classical, Gross-Pitaevskii equations. For an initial Fock state, while each trajectory is deterministic, the initial state is chosen from a distribution in the doubled phase space of the representation. The equations are as given in section~\ref{sec:stochastic} below, but without the noise terms. We present the solutions for the populations here, beginning again with an initial Fock state of $200$ atoms in the first well.

\begin{figure}
\begin{center}
\includegraphics[width=0.8\columnwidth]{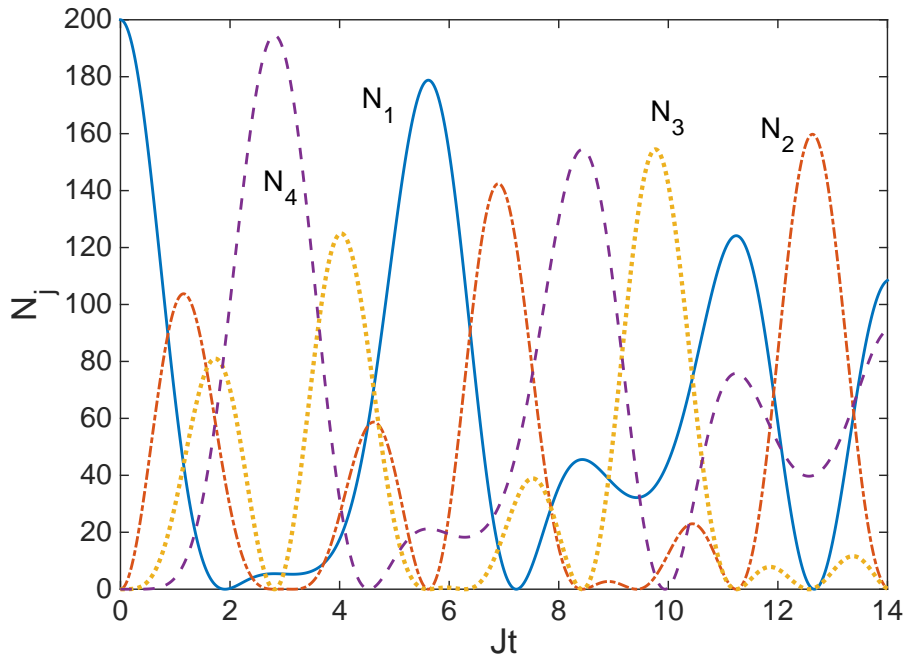}
\end{center}
\caption{(Color online) The numerical non-interacting solutions for the populations of the four-well model as a function of time, for $J=1$, and $N_{1}(0)=200$ in a Fock state, with $N_{2}(0)=N_{3}(0)= N_{4}(0)=0$.  These solutions were averaged over $2.88\times 10^{5}$ stochastic realisations of the positive-P equations. We see that the populations now exhibit much more complicated dynamics than for the three-well case shown in Fig.~\ref{fig:populations3}.}
\label{fig:pops4}
\end{figure}

\section{Analytic quantum correlations}
\label{sec:correlations}

As well as the populations in each well, we also calculate any type of operator products that we desire, analytically in the case without interactions. 
We will calculate the number variances in each well, and entanglement and EPR steering measures adapted from an inequality developed by Hillery and Zubairy. They showed that, considering two separable modes denoted by $i$ and $j$~\cite{HZ},
\begin{equation}
| \langle \hat{a}_{i}^{\dag}\hat{a}_{j}\rangle |^{2} \leq \langle \hat{a}_{i}^{\dag}\hat{a}_{i}\hat{a}_{j}^{\dag}\hat{a}_{j}\rangle,
\label{eq:HZ}
\end{equation}
with the equality holding for coherent states. The violation of this inequality is thus an indication of the inseparability of, and entanglement between, the two modes. Cavalcanti \etal\cite{ericsteer} have extended this inequality to provide indicators of EPR 
steering~\cite{Einstein,Erwin,Wiseman} and Bell violations~\cite{Bell}. We now define the correlation function
\begin{equation}
\xi_{ij} = \langle \hat{a}_{i}^{\dag}\hat{a}_{j}\rangle\langle \hat{a}_{i}\hat{a}_{j}^{\dag}\rangle - \langle \hat{a}_{i}^{\dag}\hat{a}_{i}\hat{a}_{j}^{\dag}\hat{a}_{j}\rangle,
\label{eq:xiij}
\end{equation}
for which a positive value reveals entanglement between modes $i$ and $j$. We easily see that $\xi_{ij}$ gives a value of zero for two independent coherent states and a negative result for two independent Fock states. 

Cavalcanti \etal~\cite{ericsteer} further developed the work of Hillery and Zubairy to find inequalities for which the violation denotes the possibility of EPR steering and Bell states. The EPR steering inequality for two modes is written as
\begin{equation}
|\langle \hat{a}_{i}\hat{a}_{j}^{\dag}\rangle|^{2} \leq \langle \hat{a}_{i}^{\dag}\hat{a}_{i}(\hat{a}_{j}^{\dag}\hat{a}{j}+\frac{1}{2})\rangle ,
\label{eq:ericEPR}
\end{equation}
while the Bell state inequality is written as
\begin{equation}
|\langle \hat{a}_{i}\hat{a}_{j}^{\dag}\rangle|^{2} \leq \langle (\hat{a}_{i}^{\dag}\hat{a}_{i}+\frac{1}{2})(\hat{a}_{j}^{\dag}\hat{a}{j}+\frac{1}{2})\rangle ,
\label{eq:ericBell}
\end{equation}
Calling on the overworked Alice and Bob, if Alice measures mode $i$ and Bob measures mode $j$ a violation of the inequality (\ref{eq:ericEPR}) signifies that Bob would be able to steer Alice, and vice versa for a swapping of the modes. These inequalities allow us to define a correlation function which signifies the presence of EPR steering when it has a value of greater than zero,
\begin{equation}
\Sigma_{ij} = \langle \hat{a}_{i}\hat{a}_{j}^{\dag}\rangle\langle \hat{a}_{i}^{\dag}\hat{a}_{j}\rangle - \langle \hat{a}_{i}^{\dag}\hat{a}_{i}(\hat{a}_{j}^{\dag}\hat{a}{j}+\frac{1}{2})\rangle ,
\label{eq:EPRsigma}
\end{equation}
and another for which a positive value signifies the presence of Bell correlations,
\begin{equation}
\zeta_{ij} = \langle \hat{a}_{i}\hat{a}_{j}^{\dag}\rangle\langle \hat{a}_{i}^{\dag}\hat{a}_{j}\rangle - \langle (\hat{a}_{i}^{\dag}\hat{a}_{i}+\frac{1}{2})(\hat{a}_{j}^{\dag}\hat{a}{j}+\frac{1}{2})\rangle .
\label{eq:Bellbeta}
\end{equation}
The criteria $\xi_{ij}$ and $\Sigma_{ij}$ have been shown to detect both inseparability and asymmetric EPR steering in a three-well Bose-Hubbard model under the process of coherent transfer of atomic population 
(CTAP)~\cite{myJPB,myJOSAB}, as well as bipartite entanglement in a three-mode model with all population initially in the central well~\cite{toberejected,BS3expand}. 

\subsection{Twin well model}
\label{subsec:twinHZ}

\begin{figure}
\begin{center}
\includegraphics[width=0.8\columnwidth]{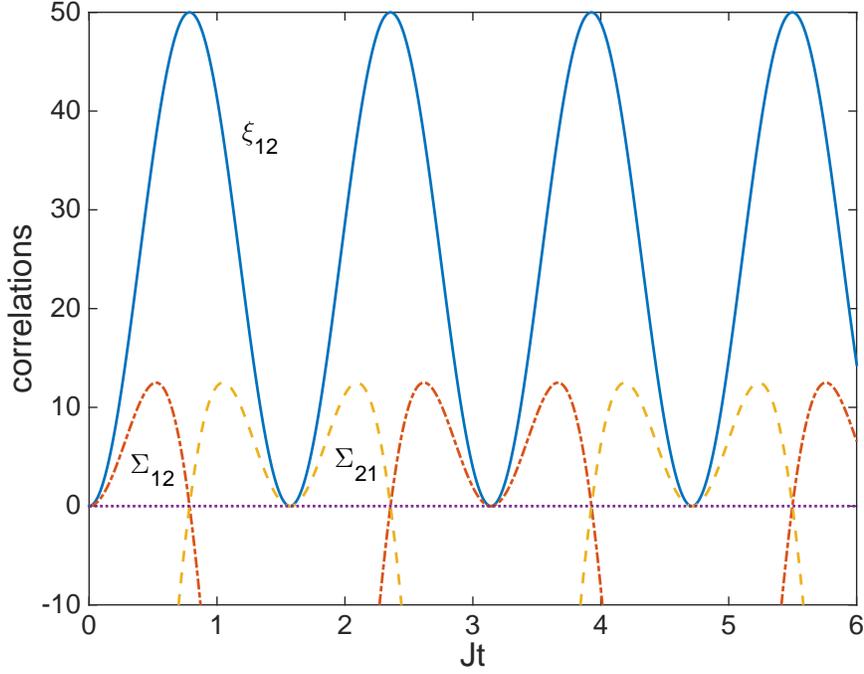}
\end{center}
\caption{(Color online) The analytical non-interacting solutions for $\xi_{12}$ and $\Sigma_{ij}$ in the twin-well model as a function of time, for $J=1$ and $N_{1}(0)=200$ in a Fock state, with $N_{2}(0)=N_{3}(0)=0$. We see that the $\Sigma_{ij}$ exhibit positivity at different times, showing that asymmetric steering is possible by this measure.}
\label{fig:bicorrs}
\end{figure}

The first class of correlations we calculate are the number variances in each well. In terms of the operators, these are 
\begin{equation}
V(\hat{N}_{j}) = \langle \hat{a}_{j}^{\dag}\hat{a}_{j}\hat{a}_{j}^{\dag}\hat{a}_{j}\rangle - \langle \hat{a}_{j}^{\dag}\hat{a}_{j}\rangle^{2}.
\label{eq:VNernest}
\end{equation}
We find, dropping the time argument for simplicity,
\begin{eqnarray}
V(\hat{N}_{1}) &=& V(\hat{N}_{1}(0))\cos^{4}Jt + \frac{1}{4}\langle\hat{N}_{1}(0)\rangle \sin^{2}2Jt,\nonumber\\
V(\hat{N}_{2}) &=& V(\hat{N}_{1}(0))\sin^{4}Jt + \frac{1}{4}\langle\hat{N}_{1}(0)\rangle \sin^{2}2Jt.
\label{eq:VN2}
\end{eqnarray}
The Hillery-Zubairy correlation is found as
\begin{equation}
\xi_{12} = \frac{1}{4}\sin^{2}2Jt\left[\langle \hat{N}_{1}(0)\rangle - V(\hat{N}_{1}(0)) \right],
\label{eq:HZ2}
\end{equation}
showing that entanglement will be detected by this measure whenever the initial state of the atoms in the first well is sub-Poissonian. The correlation will be maximised for an intial Fock state, as shown in Fig.~\ref{fig:bicorrs}. 

The EPR steering correlations are solved as
\begin{eqnarray}
\Sigma_{12} &=& \frac{1}{4}\sin^{2}2Jt\left[\langle\hat{N}_{1}(0)\rangle-V(\hat{N}_{1}(0))\right]-\frac{1}{2}\sin^{2}Jt\langle\hat{N}_{1}(0)\rangle, \nonumber\\
\Sigma_{21} &=& \frac{1}{4}\sin^{2}2Jt\left[\langle\hat{N}_{1}(0)\rangle-V(\hat{N}_{1}(0))\right]-\frac{1}{2}\cos^{2}Jt\langle\hat{N}_{1}(0)\rangle,
\label{eq:twinsigma}
\end{eqnarray}
both of which can be positive. We note that $\Sigma_{12}\neq\Sigma_{21}$, which leaves open the possibility of asymmetric steering~\cite{asymSarah}, which we see in Fig.~\ref{fig:bicorrs}, where $\Sigma_{12}$ and $\Sigma_{21}$ are positive at different times for an initial Fock state. We note here that, although the steering is asymmetric by this measure, this does not mean this is the case for all possible measures.
The Bell correlation is found as
\begin{equation}
\zeta_{12} = \frac{1}{4}\sin^{2}2Jt\left[\langle \hat{N}_{1}(0)\rangle - V(\hat{N}_{1}(0))\right] - \langle\hat{N}_{1}\rangle-\frac{1}{4},
\label{eq:Bell2}
\end{equation}
which obviously can never be positive.

\subsection{Triple well model}
\label{subsec:tripleHZ}

\begin{figure}
\begin{center}
\includegraphics[width=0.8\columnwidth]{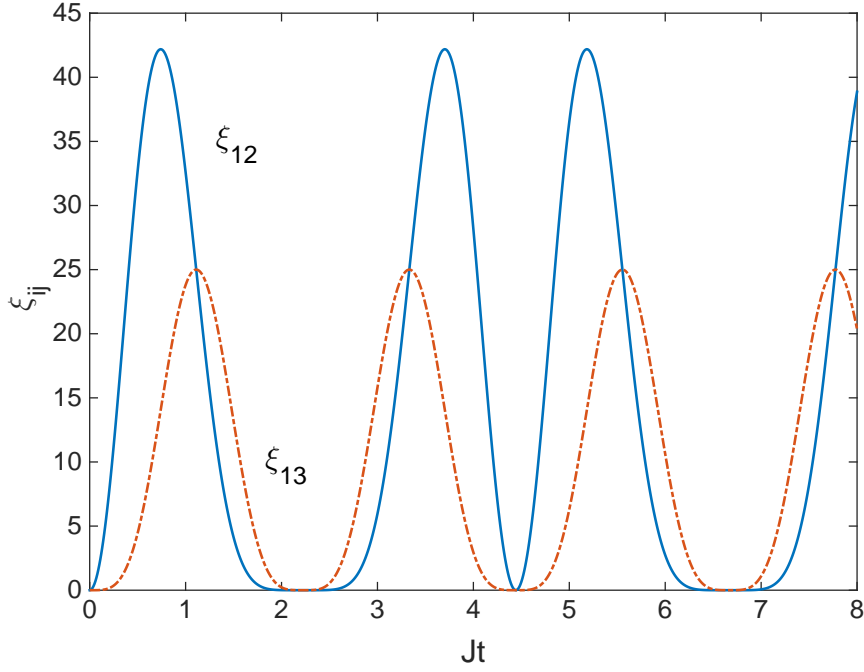}
\end{center}
\caption{(Color online) The analytical non-interacting solutions for $\xi_{12}$ and $\xi_{13}$ in the triple-well model as a function of time, for $J=1$ and $N_{1}(0)=200$ in a Fock state, with $N_{2}(0)=N_{3}(0)=0$. We see that the maximum of $\xi_{12}$ is greater than the maximum of $\xi_{13}$, suggesting that entanglement decreases with distance along the chain.}
\label{fig:HZ3}
\end{figure}

In the non-interacting case and again with only the first well initially occupied, we find
\begin{eqnarray}
V(\hat{N}_{1}) &=& \frac{1}{16}\left(1+\cos\Omega t\right)^{4}V(\hat{N}_{1}(0)) + \frac{1}{8}\sin^{2}\Omega t\left[\frac{1}{2}\sin^{2}\Omega t + (1+\cos\Omega t)^{2}\right]\langle\hat{N}_{1}(0)\rangle, \nonumber \\
V(\hat{N}_{2}) &=& \frac{1}{4}\sin^{4}\Omega t\;V(\hat{N}_{1}(0))+\frac{1}{2}\sin^{2}\Omega t\left(1-\frac{1}{2}\sin^{2}\Omega t\right)\langle \hat{N}_{1}(0)\rangle, \nonumber \\
V(\hat{N}_{3}) &=& \frac{1}{16}\left(\cos\Omega t-1 \right)^{4}V(\hat{N}_{1}(0))+\frac{1}{4}\left(\cos\Omega t-1\right)^{2}\langle\hat{N}_{1}(0)\rangle.
\label{eq:Heisvars3}
\end{eqnarray}
The Hillery-Zubairy inseparability correlations $\xi_{1j}$ are found as
\begin{eqnarray}
\xi_{12} &=& \frac{1}{8}\sin^{2}\Omega t\left(1+\cos\Omega t \right)^{2}\left[\langle\hat{N}_{1}(0)\rangle -V(\hat{N}_{1}(0)) \right], \nonumber \\
\xi_{13} &=& \frac{1}{16}\sin^{4}\Omega t\left[\langle \hat{N}_{1}(0)-V(\hat{N}_{1}(0)) \right],
\label{eq:HZtriple}
\end{eqnarray}
from which we can see that both modes $1$ and $2$, and $1$ and $3$ can exhibit a time dependent inseparability for an initial sub-Poissonian state in the first well. We also see from both the equations and Fig.~\ref{fig:HZ3}  that the maximum value of the function is less for $\xi_{13}$ than for $\xi_{12}$, indicating that the entanglement with the first well decreases with distance along the chain.

\begin{figure}
\begin{center}
\includegraphics[width=0.8\columnwidth]{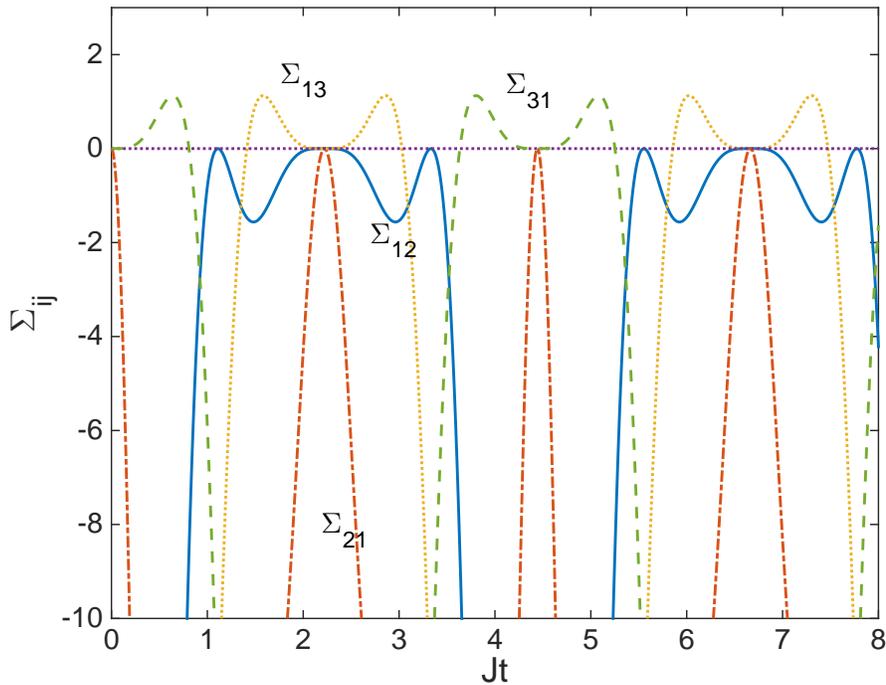}
\end{center}
\caption{(Color online) The analytical non-interacting solutions for the $\Sigma_{ij}$ in the triple-well model as a function of time, for $J=1$ and $N_{1}(0)=200$ in a Fock state, with $N_{2}(0)=N_{3}(0)=0$. We see that only $\Sigma_{13}$ and $\Sigma_{31}$ take on positive values.}
\label{fig:EPR3}
\end{figure}

For the EPR steering criteria we find
\begin{eqnarray}
\Sigma_{12} &=& -\frac{1}{8}\left(1+\cos\Omega t\right)^{2}\left[ \langle\hat{N}_{1}(0)\rangle \cos^{2}\Omega t+V(\hat{N}_{1}(0)) \right], 
\nonumber \\
\Sigma_{21} &=& -\frac{1}{8}\sin^{2}\Omega t\left\{\left(1+\cos\Omega t\right)^{2}\left[V(\hat{N}_{1}(0))+\langle\hat{N}_{1}(0)\rangle\right] +2\cos^{2}\Omega t\langle\hat{N}_{1}(0)\rangle\right\}, \nonumber \\
\Sigma_{13} &=& \frac{1}{16}\sin^{4}\Omega t\left[\langle\hat{N}_{1}(0)\rangle - V(\hat{N}_{1}(0))\right] - \frac{1}{8}\left(1+\cos\Omega t\right)^{2}\langle\hat{N}_{1}(0)\rangle,   \nonumber \\
\Sigma_{31} &=& \frac{1}{16} \sin^{4}\Omega t\left[\langle\hat{N}_{1}(0)\rangle-V(\hat{N}_{1}(0))\right]-\frac{1}{8}\left(\cos\Omega t-1\right)^{2}\langle\hat{N}_{1}(0)\rangle,
\label{eq:EPR3}
\end{eqnarray}
none of which can be greater than zero for initial coherent states. For initial Fock states, we find that $\Sigma_{13}$ and $\Sigma_{31}$ can take on positive values, and we again see that the correlations measured are not equivalent under an exchange of indices. These are shown in Fig.~\ref{fig:EPR3}.

\section{Stochastic methods}
\label{sec:stochastic}

In this case ($\chi\neq 0$), it is not obvious how to solve the full quantum equations of motion analytically.
We will therefore use the positive-P representation~\cite{Pplus}, which allows for exact solutions of the dynamics arising from the Hamiltonians, in the limit of the average of an infinite number of trajectories of the stochastic differential equations in a doubled phase-space. In practice we obviously cannot integrate an infinite number of trajectories, but have used numbers large enough that the sampling error is within the line thicknesses of our plotted results.
Following the standard methods~\cite{DFW}, the set of It\^o stochastic differential equations~\cite{SMCrispin} for the two-well system are found as
\begin{eqnarray}
\frac{d\alpha_{1}}{dt} &=& -2i\chi\alpha_{1}^{+}\alpha_{1}^{2}-iJ\alpha_{2}
+\sqrt{-2i\chi\alpha_{1}^{2}}\;\eta_{1},\nonumber\\
\frac{d\alpha_{1}^{+}}{dt} &=& 2i\chi\alpha_{1}^{+\,2}\alpha_{1}+iJ\alpha_{2}^{+}
+\sqrt{2i\chi\alpha_{1}^{+\;2}}\;\eta_{2},\nonumber\\
\frac{d\alpha_{2}}{dt} &=& -2i\chi\alpha_{2}^{+}\alpha_{2}^{2}-iJ\alpha_{1}
+\sqrt{-2i\chi\alpha_{2}^{2}}\;\eta_{3},\nonumber\\
\frac{d\alpha_{2}^{+}}{dt} &=& 2i\chi\alpha_{2}^{+\,2}\alpha_{2} + iJ\alpha_{1}^{+}
+\sqrt{2i\chi\alpha_{2}^{+\,2}}\;\eta_{4},\nonumber\\
\label{eq:Pplus2}
\end{eqnarray}
where the $\eta_{j}$ are standard Gaussian noises with $\overline{\eta_{j}}=0$ and $\overline{\eta_{j}(t)\eta_{k}(t')}=\delta_{jk}\delta(t-t')$. As always, averages of the positive-P variables represent normally ordered operator moments, such that, for example, $\overline{\alpha_{j}^{m}\alpha_{k}^{+\,n}}\rightarrow\langle\hat{a}^{\dag\,n}\hat{a}^{m}\rangle$. We also note that $\alpha_{j}=(\alpha_{j}^{+})^{\ast}$ only after taking averages, and it is this freedom that allows classical variables to represent quantum operators.

The equations for the triple and quadruple well systems are obvious extensions of Eq.~\ref{eq:Pplus2}, but we will present them here for the sake of completeness. For the three-well system we find
\begin{eqnarray}
\frac{d\alpha_{1}}{dt} &=& -2i\chi\alpha_{1}^{+}\alpha_{1}^{2}-iJ\alpha_{2}
+\sqrt{-2i\chi\alpha_{1}^{2}}\;\eta_{1},\nonumber\\
\frac{d\alpha_{1}^{+}}{dt} &=& 2i\chi\alpha_{1}^{+\,2}\alpha_{1}+iJ\alpha_{2}^{+}
+\sqrt{2i\chi\alpha_{1}^{+\;2}}\;\eta_{2},\nonumber\\
\frac{d\alpha_{2}}{dt} &=& -2i\chi\alpha_{2}^{+}\alpha_{2}^{2}-iJ\left(\alpha_{1}
+\alpha_{3}\right)
+\sqrt{-2i\chi\alpha_{2}^{2}}\;\eta_{3},\nonumber\\
\frac{d\alpha_{2}^{+}}{dt} &=& 2i\chi\alpha_{2}^{+\,2}\alpha_{2} + iJ\left(\alpha_{1}^{+}
+\alpha_{3}^{+}\right)
+\sqrt{2i\chi\alpha_{2}^{+\,2}}\;\eta_{4},\nonumber\\
\frac{d\alpha_{3}}{dt} &=& -2i\chi\alpha_{3}^{+}\alpha_{3}^{2}-iJ\alpha_{2}
+\sqrt{-2i\chi\alpha_{3}^{2}}\;\eta_{5},\nonumber\\
\frac{d\alpha_{3}^{+}}{dt} &=& 2i\chi\alpha_{3}^{+\,2}\alpha_{3} + iJ\alpha_{2}^{+}
+\sqrt{2i\chi\alpha_{3}^{+\;2}}\;\eta_{6},
\label{eq:Pplus3}
\end{eqnarray}
while the equations for the four-well system are
\begin{eqnarray}
\frac{d\alpha_{1}}{dt} &=& -2i\chi\alpha_{1}^{+}\alpha_{1}^{2}-iJ\alpha_{2}
+\sqrt{-2i\chi\alpha_{1}^{2}}\;\eta_{1},\nonumber\\
\frac{d\alpha_{1}^{+}}{dt} &=& 2i\chi\alpha_{1}^{+\,2}\alpha_{1}+iJ\alpha_{2}^{+}
+\sqrt{2i\chi\alpha_{1}^{+\;2}}\;\eta_{2},\nonumber\\
\frac{d\alpha_{2}}{dt} &=& -2i\chi\alpha_{2}^{+}\alpha_{2}^{2}-iJ\left(\alpha_{1}
+\alpha_{3}\right)
+\sqrt{-2i\chi\alpha_{2}^{2}}\;\eta_{3},\nonumber\\
\frac{d\alpha_{2}^{+}}{dt} &=& 2i\chi\alpha_{2}^{+\,2}\alpha_{2} + iJ\left(\alpha_{1}^{+}
+\alpha_{3}^{+}\right)
+\sqrt{2i\chi\alpha_{2}^{+\,2}}\;\eta_{4},\nonumber\\
\frac{d\alpha_{3}}{dt} &=& -2i\chi\alpha_{3}^{+}\alpha_{3}^{2}-iJ\left(\alpha_{2}
+\alpha_{4}\right)
+\sqrt{-2i\chi\alpha_{3}^{2}}\;\eta_{5},\nonumber\\
\frac{d\alpha_{3}^{+}}{dt} &=& 2i\chi\alpha_{3}^{+\,2}\alpha_{3} + iJ\left(\alpha_{2}^{+}
+\alpha_{4}^{+}\right)
+\sqrt{2i\chi\alpha_{3}^{+\,2}}\;\eta_{6},\nonumber\\
\frac{d\alpha_{4}}{dt} &=& -2i\chi\alpha_{4}^{+}\alpha_{4}^{2}-iJ\alpha_{3}
+\sqrt{-2i\chi\alpha_{4}^{2}}\;\eta_{7},\nonumber\\
\frac{d\alpha_{4}^{+}}{dt} &=& 2i\chi\alpha_{4}^{+\,2}\alpha_{4} + iJ\alpha_{3}^{+}
+\sqrt{2i\chi\alpha_{4}^{+\;2}}\;\eta_{8}.
\label{eq:Pplus4}
\end{eqnarray}
These systems of equations must be solved numerically, for which we use Matlab. This allows us to average over a sufficient number of trajectories in a reasonable time, in most cases less than $2$ hours. If we wished to add wells, the growth in computational time would be linear in the number, and we found that $4$ wells did not extend our computational resources to any significant extent.

\section{Numerical solutions}
\label{sec:numerics}

For our results in the interacting case, we have chosen a nonlinearity of $\chi=10^{-3}$, again with either a Fock or coherent state with an average of $200$ atoms in the first well and all the others initially empty. These different quantum states are simulated using the methods found in Olsen and Bradley~\cite{states}. We have chosen coherent states because they are the initial state most often used in numerical simulations, and Fock states because we consider this the most natural state for atoms in an isolated well.

 \subsection{Two wells} 
 \label{subsec:PPtwin}
 
\begin{figure}
\begin{center}
\includegraphics[width=0.8\columnwidth]{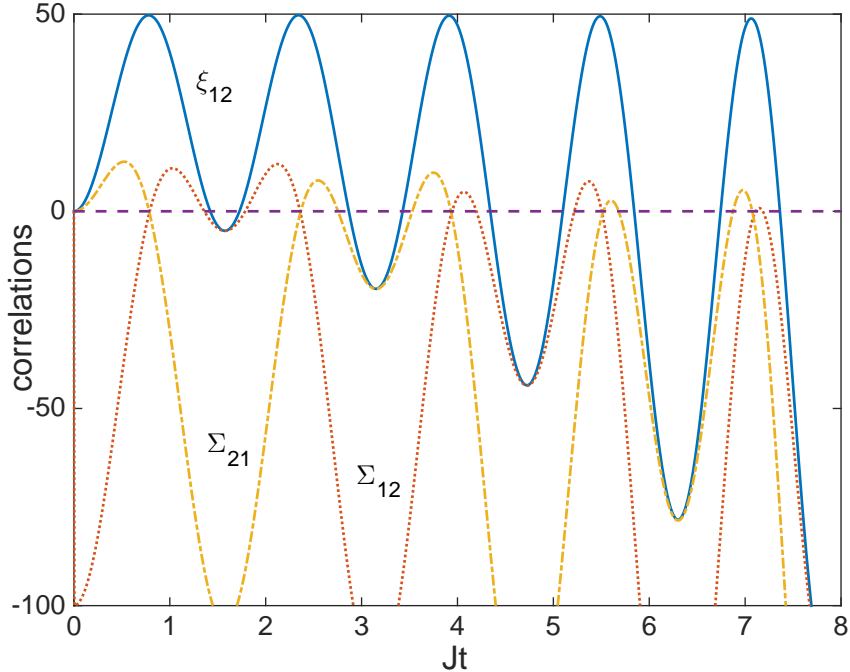}
\end{center}
\caption{(Color online) The numerical interacting solutions for $\xi_{12}$, $\Sigma_{12}$ and $\Sigma_{21}$ in the twin-well model as a function of time, for $J=1$, $\chi=10^{-3}$ and $N_{1}(0)=200$ in a Fock state, with $N_{2}(0)=0$. We see that the correlations are no longer periodic as in the non-interacting case of Fig.~\ref{fig:bicorrs}, and that the two EPR steering correlations give positive signals at different times. These solutions are averaged over $1.9\times 10^{6}$ stochastic realisations of the positive-P equations.}
\label{fig:HZPP2}
\end{figure}

For the two-well system we found evidence of both entanglement and EPR steering in the non-interacting case. We find that the results with a collisional nonlinearity and an initial Fock state are initially similar, as shown in Fig.~\ref{fig:HZPP2}, but that the correlations degrade with time. For an initial coherent state in the first well, we found no evidence of EPR steering, and only a very small positive value of $\xi_{12}\approx 0.2$ at $Jt\approx 2.2$. We note that our results are not inconsistent with those of He \etal~\cite{He}, who found entanglement in the ground state of a two-well system. Our system, beginning with only one of the wells initially occupied, effectively undergoes a sudden parameter change from $J=0$ to $J=1$ at $t=0$, and in that respect has more in common with the work of Lauchli and Kollath~\cite{quench}, who studied the time evolution of entanglement following a quench. In our case, the sudden change in parameters is the turning on of the tunnelling at $t=0$, so that our system is then far from the equilibrium state of the multi-well systems.

\subsection{Three wells}
\label{subsec:PPtri}

\begin{figure}
\begin{center}
\includegraphics[width=0.8\columnwidth]{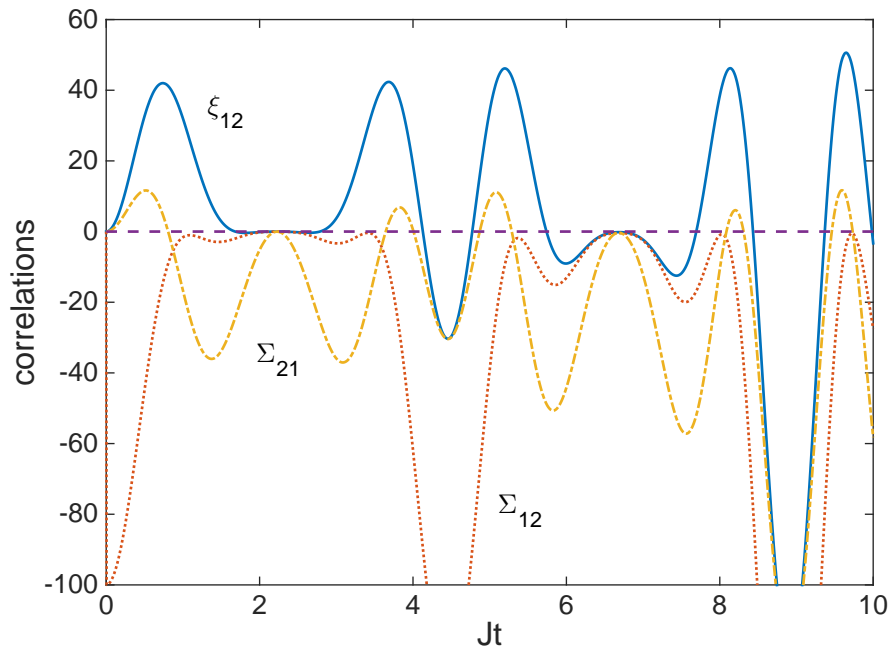}
\end{center}
\caption{(Color online) The numerical interacting solutions for $\xi_{12}$, $\Sigma_{12}$ and $\Sigma_{21}$ in the three-well model as a function of time, for $J=1$, $\chi=10^{-3}$, and $N_{1}(0)=200$ in a Fock state, with $N_{2}(0)=N_{3}(0)=0$. We see that the correlations degrade with time when compared to the non-interacting case of Fig.~\ref{fig:HZ3}, and that only one of the EPR steering correlations gives a positive signal for any of the times considered. These solutions were averaged over $6.74\times 10^{5}$ stochastic realisations of the positive-P equations.}
\label{fig:HZPP3(12)}
\end{figure}

\begin{figure}
\begin{center}
\includegraphics[width=0.8\columnwidth]{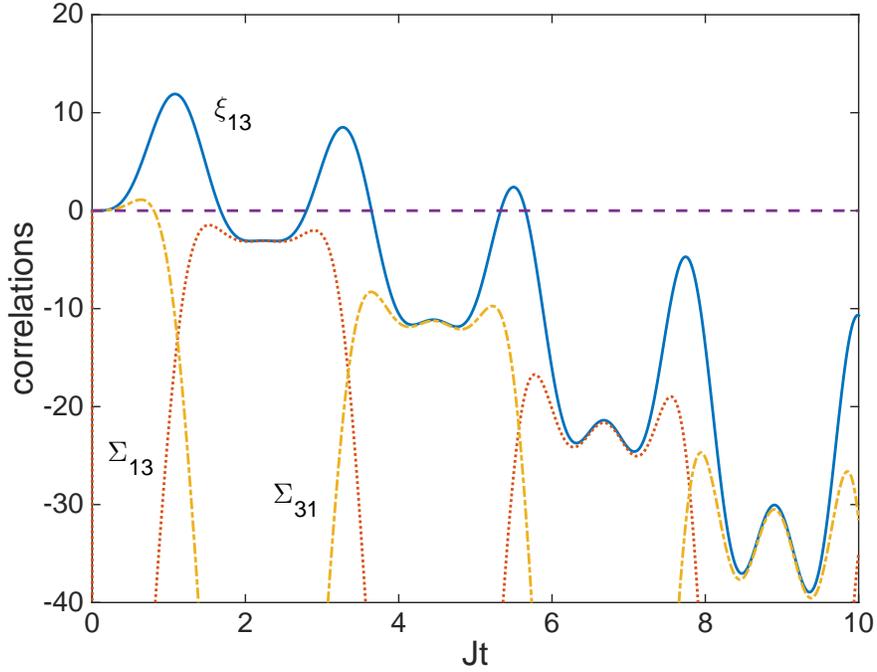}
\end{center}
\caption{(Color online) The numerical interacting solutions for $\xi_{13}$, $\Sigma_{13}$ and $\Sigma_{31}$ in the three-well model as a function of time, for $J=1$, $\chi=10^{-3}$ and $N_{1}(0)=200$ in a Fock state, with $N_{2}(0)=N_{3}(0)=0$. We see that the correlations degrade with time when compared to the non-interacting case of Fig.~\ref{fig:HZ3}, and that only $\Sigma_{31}$ gives a positive EPR steering signal, and this only for a short time. These solutions were averaged over $6.74\times 10^{5}$ stochastic realisations of the positive-P equations.}
\label{fig:HZPP3(13)}
\end{figure}

With an initial Fock state in the first well, we find signals of entanglement with both $\xi_{12}$ and $\xi_{13}$, as shown in Fig.~\ref{fig:HZPP3(12)} and Fig.~\ref{fig:HZPP3(13)}. These initially follow the analytical non-interacting solutions, but again become different with time. We find asymmetric EPR steering in both cases, with only $\Sigma_{21}$ and $\Sigma_{31}$ attaining positive values, while $\Sigma_{12}$ and $\Sigma_{13}$ remain negative.

For an initial coherent state, we see no evidence of EPR steering in the three-well model, and only a very small signal of entanglement between wells $1$ and $2$, with $\xi_{12}\leq 2.5$, and even that only for very short intervals of time. $\xi_{13}$ had a maximum value of zero over the length of time of our investigations.

\subsection{Four wells}
\label{subsec:PPquad}

The solutions for the populations of the four-well system become more irregular, without the clear periodicity of the smaller chains, as was shown above in Fig.~\ref{fig:pops4} for an initial Fock state in the non-interacting case. The solutions when we begin with either Fock or coherent states are indistinguishable over the time interval investigated. When we add interactions, the solutions begin to deviate from those without interactions around $Jt=4$, with the deviations increasing with time. This is common with Bose-Hubbard models and has been seen previously in two-, three-, and four well systems~\cite{BHJoel,Chiancathermal}. The irregularity and lack of clear periodicity may be a sign of the onset of chaos, which has been predicted in longer chains, but with much smaller well occupation numbers~\cite{Kolovsky,Carleo}. Irregular behaviour has previously been seen in a four-well system in a square configuration with two different tunnelling rates~\cite{Chianca4,Chiancathermal}, and may happen because we begin with the system far from equilibrium. We integrated the classical equations as far as $t=60$ and the solutions continued to be highly irregular. However, further investigation is beyond the scope of the present work.

\begin{figure}
\begin{center}
\includegraphics[width=0.8\columnwidth]{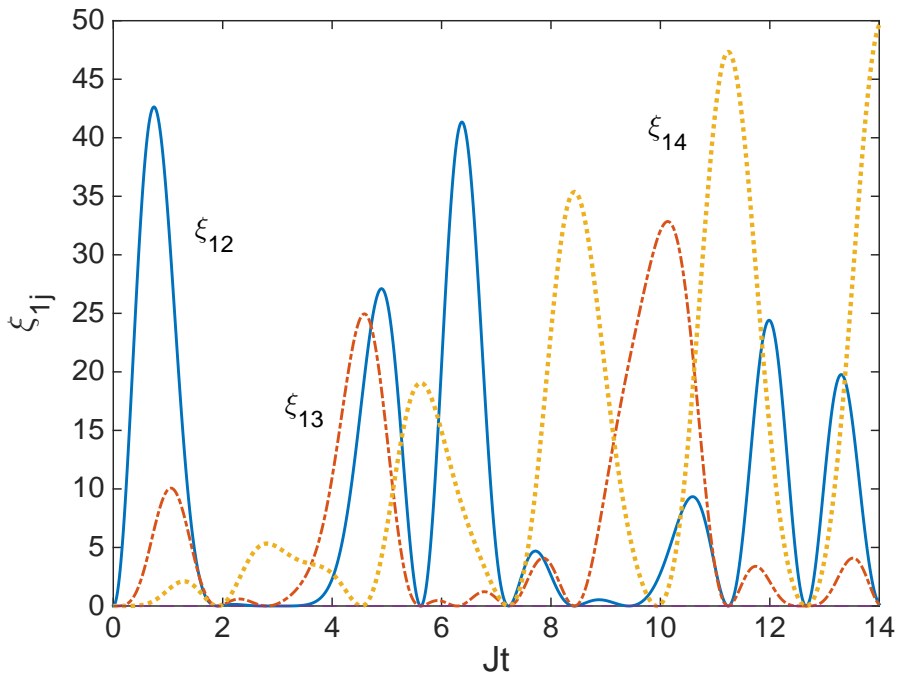}
\end{center}
\caption{(Color online) The numerical non-interacting solutions for the Hillery-Zubairy criteria of the four-well model as a function of time, for $J=1$, and $N_{1}(0)=200$ in a Fock state, with $N_{2}(0)=N_{3}(0)= N_{4}(0)=0$.  These solutions were averaged over $2.88\times 10^{5}$ stochastic realisations of the positive-P equations.}
\label{fig:HZ4xiZ}
\end{figure}

\begin{figure}
\begin{center}
\includegraphics[width=0.8\columnwidth]{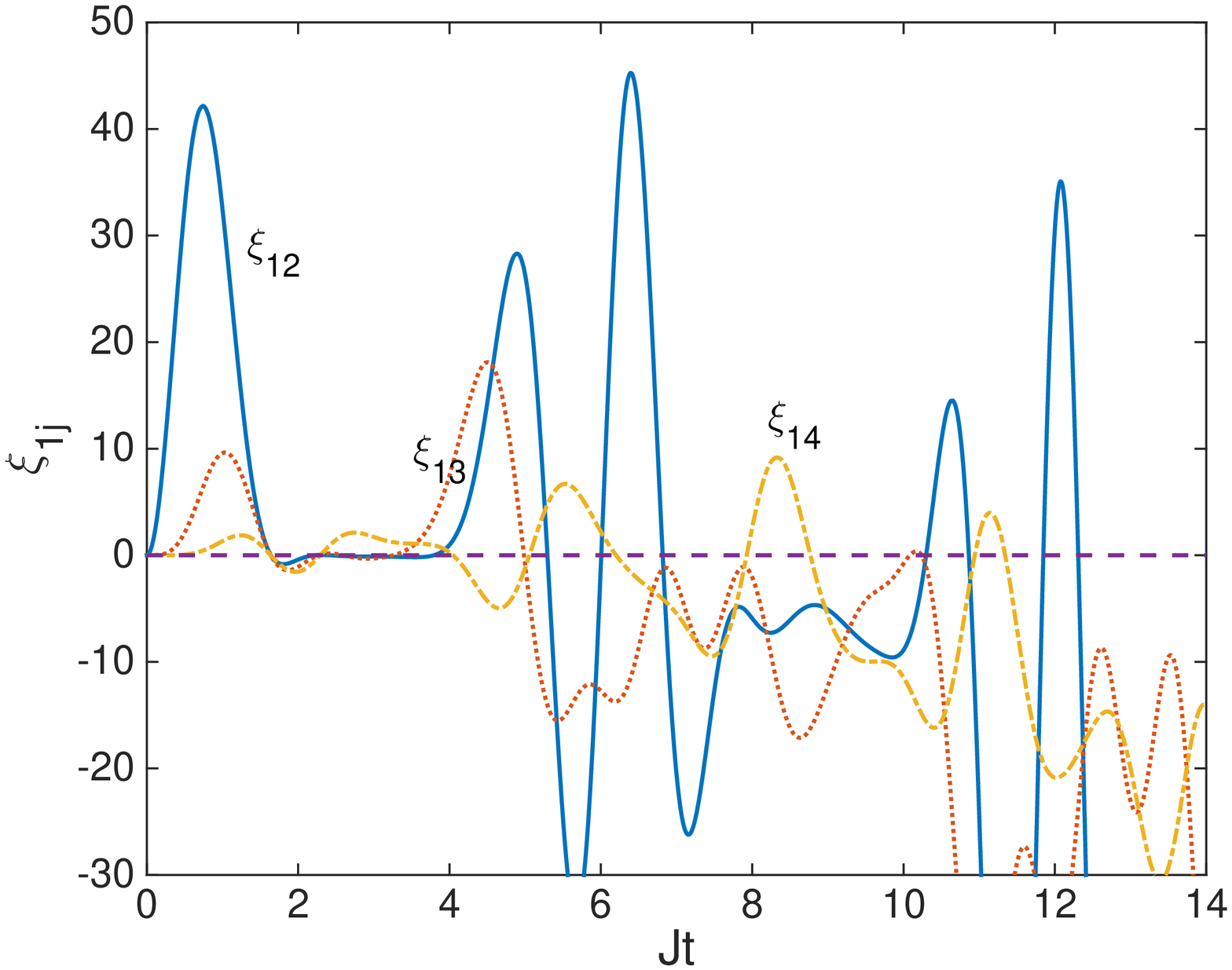}
\end{center}
\caption{(Color online) The numerical interacting solutions for the Hillery-Zubairy criteria of the four-well model as a function of time, for $J=1$, $\chi=10^{-3}$ and $N_{1}(0)=200$ in a Fock state, with $N_{2}(0)=N_{3}(0)= N_{4}(0)=0$.  These solutions were averaged over $5.54\times 10^{5}$ stochastic realisations of the positive-P equations.}
\label{fig:HZ4xi}
\end{figure}

In Fig.~\ref{fig:HZ4xiZ} we show the non-interacting $\xi_{1j}$ values with an initial Fock state, averaged over $2.88\times 10^{5}$ trajectories. We see that all values are positive semi-definite and irregular in time. The same correlations with $\chi=10^{-3}$ are presented in Fig.~\ref{fig:HZ4xi}. 
These were averaged over $5.54\times 10^{5}$ trajectories, and show clearly how interactions degrade the entanglement that can be detected by the Hillery-Zubairy measure.  As expected, we see that this measure shows the onset of entanglement between the first well and the others at times which increase for the position along the chain. The entanglement between wells one and four with interactions is also seen to persist at later times than that between one and three, at least over this time interval. When we begin with an initial coherent state in the first well, only $\xi_{12}$ takes on a positive value over the evolution time, with $\xi_{12}\approx 1$ around $Jt=6.35$. 

\begin{figure}
\begin{center}
\includegraphics[width=0.8\columnwidth]{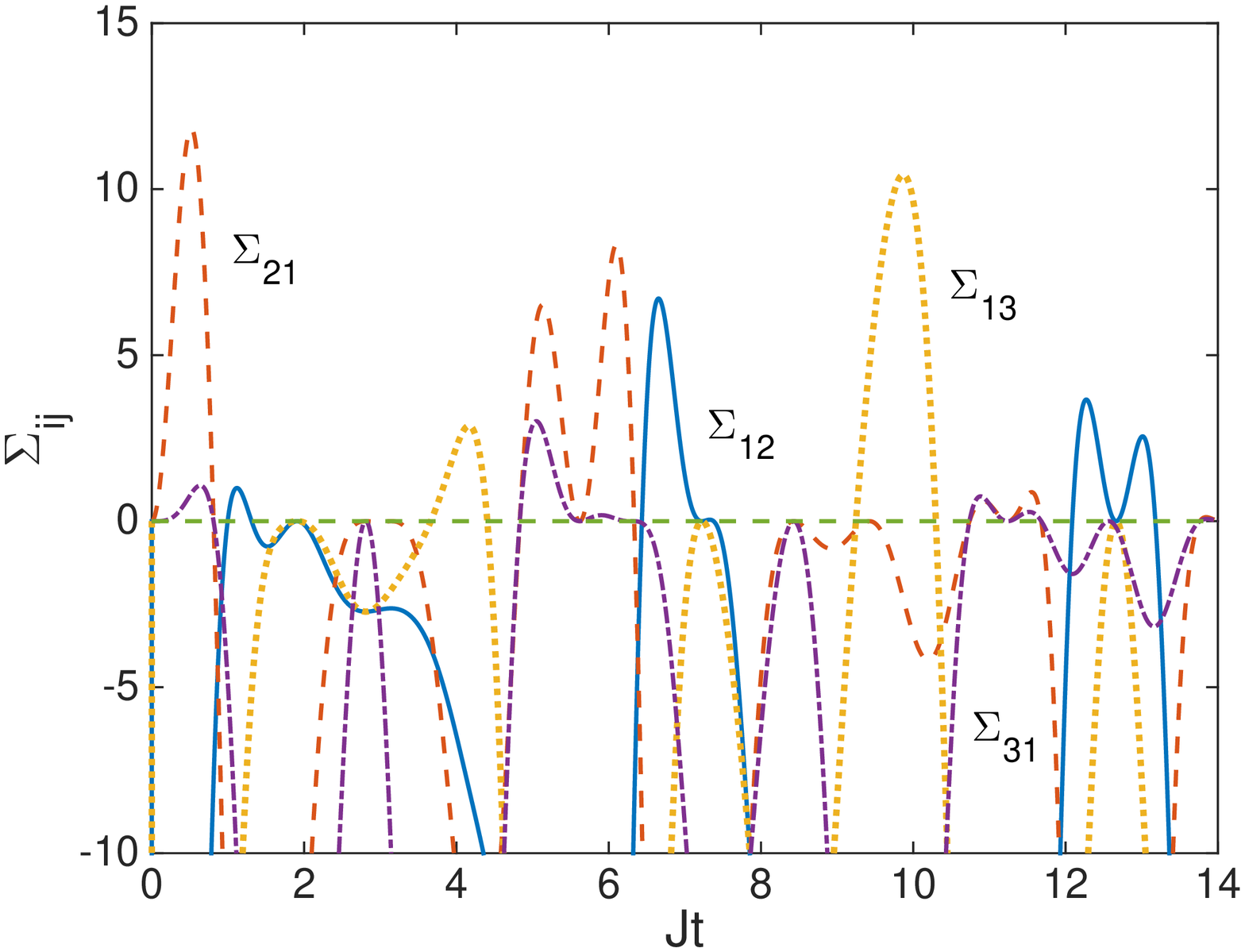}
\end{center}
\caption{(Color online) The numerical non-interacting solutions for the EPR steering criteria of the four-well model as a function of time, for $J=1$ and $N_{1}(0)=200$ in a Fock state, with $N_{2}(0)=N_{3}(0)= N_{4}(0)=0$.  These solutions were averaged over $2.88\times 10^{5}$ stochastic realisations of the positive-P equations. We again see that the system is asymmetric with regard to these measures of EPR steering.}
\label{fig:EPR123Z}
\end{figure}

\begin{figure}
\begin{center}
\includegraphics[width=0.8\columnwidth]{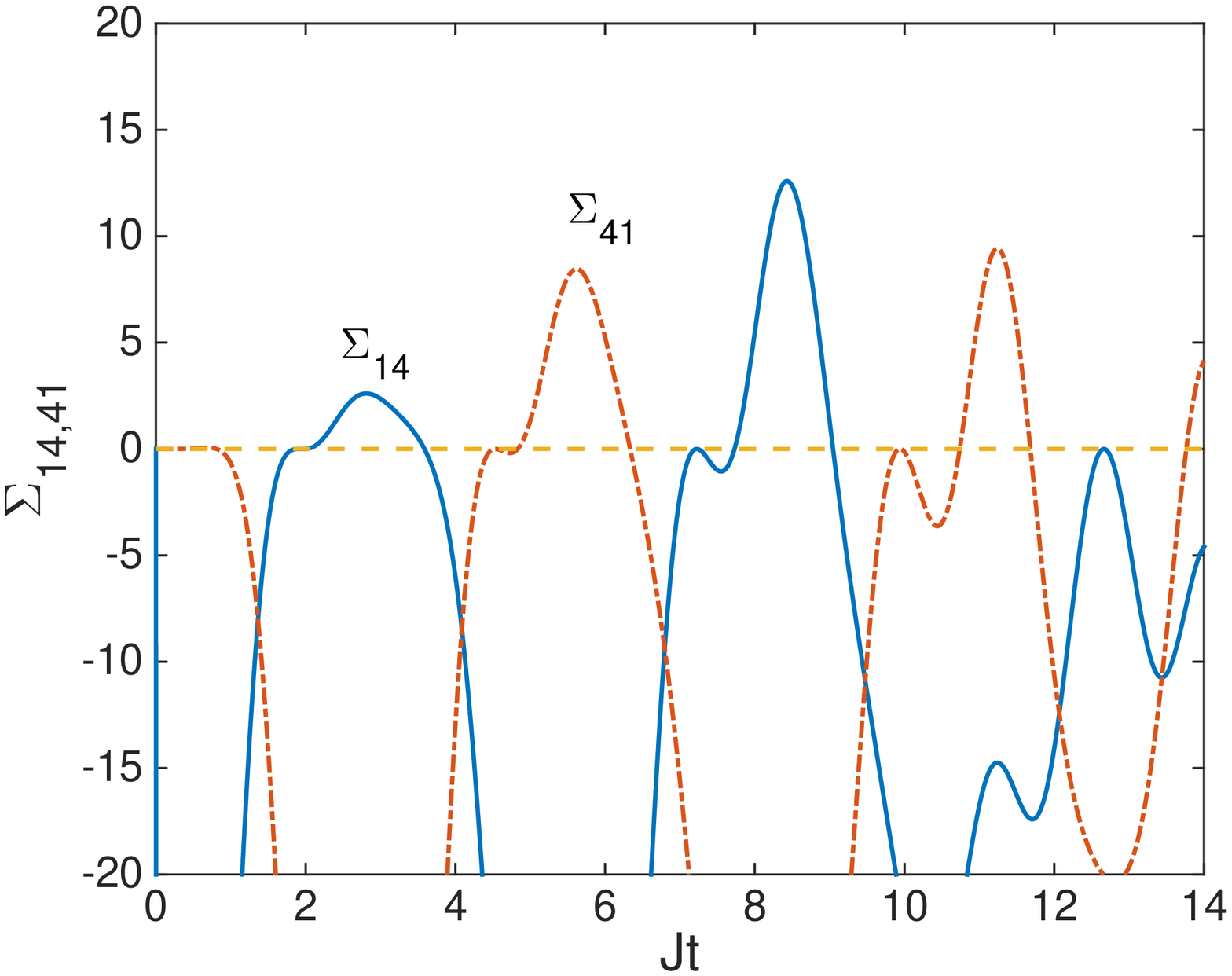}
\end{center}
\caption{(Color online) The numerical non-interacting solutions for the EPR steering criteria $\Sigma_{14}$ and $\Sigma_{41}$, for $J=1$, $\chi=10^{-3}$ and $N_{1}(0)=200$ in a Fock state, with $N_{2}(0)=N_{3}(0)= N_{4}(0)=0$.  These solutions were averaged over $2.88\times 10^{5}$ stochastic realisations of the positive-P equations.}
\label{fig:EPR4Z}
\end{figure}

There are six different bipartite EPR steering correlations between the first well and the other three. In the non-interacting case with an initial Fock state, they all take on positive values at some times, as shown in Fig.~\ref{fig:EPR123Z} and Fig.~\ref{fig:EPR4Z}. As found with the two-and three-well systems, $\Sigma_{ij}$ and $\Sigma_{ji}$ were never found to be positive at the same time. This hints that asymmetric EPR steering may not be at all uncommon in nature, although it is not possible to make a definitive claim using only one type of measurement.

\begin{figure}
\begin{center}
\includegraphics[width=0.8\columnwidth]{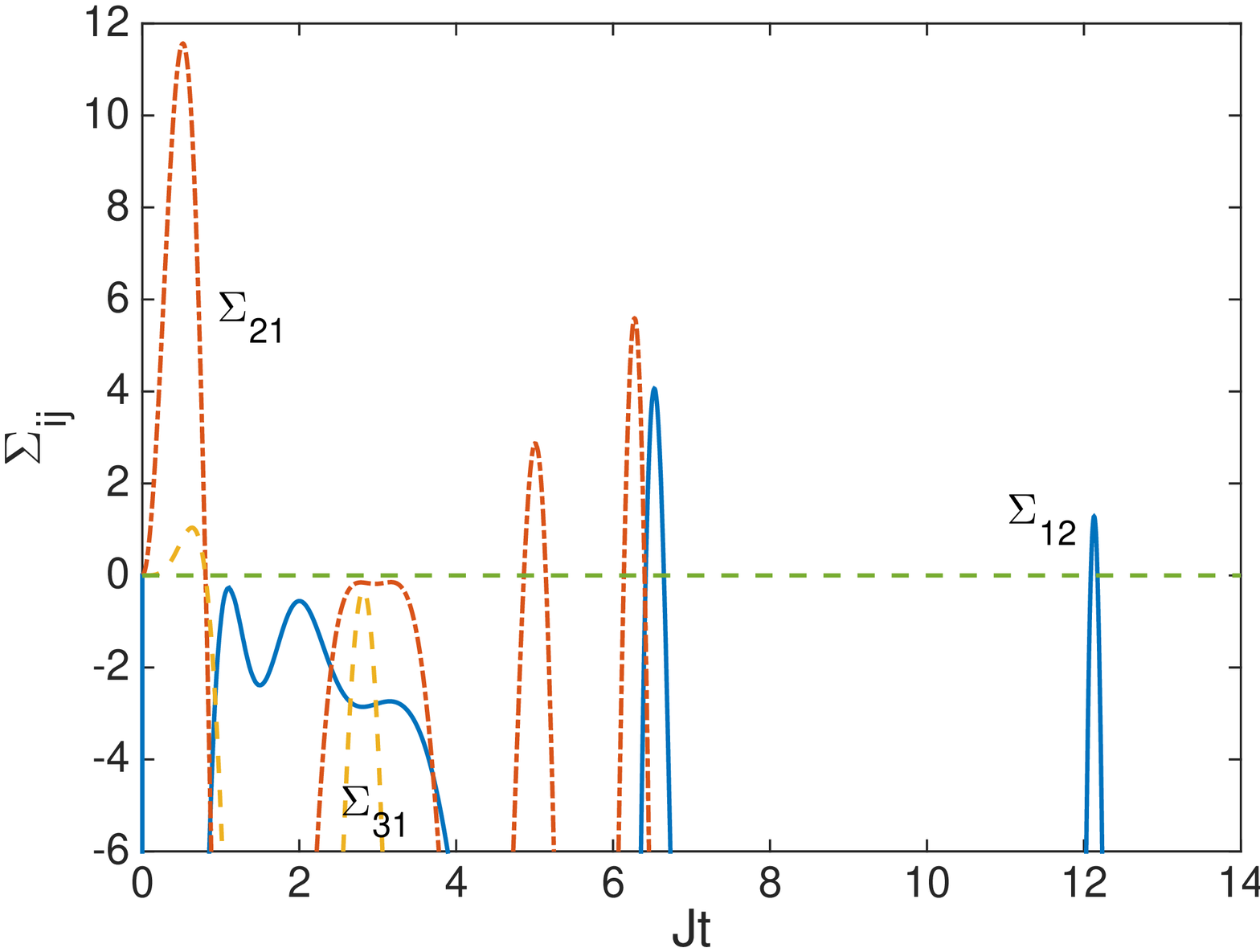}
\end{center}
\caption{(Color online) The numerical interacting solutions for the EPR steering criteria of the four-well model as a function of time, for $J=1$, $\chi=10^{-3}$ and $N_{1}(0)=200$ in a Fock state, with $N_{2}(0)=N_{3}(0)= N_{4}(0)=0$.  These solutions were averaged over $5.54\times 10^{5}$ stochastic realisations of the positive-P equations. Apart from $\Sigma_{41}$, which momentarily attained a value of less than $0.05$ around $Jt=0.7$, those not shown remained negative.}
\label{fig:EPR4}
\end{figure}

The results for the EPR steering correlations for the interacting system with an initial Fock state are presented in Fig.~\ref{fig:EPR4}. Out of the six possible bipartite correlations, only the three shown are ever significantly positive. $\Sigma_{41}$ momentarily attained a value of less than $0.05$ around $Jt=0.7$, which would be unlikely to be of practical importance. With an initial coherent state, these EPR steering measures did not take on positive values over the time interval investigated. This demonstrates that the interactions mainly serve to degrade the correlations and that, if we wish to see entanglement and EPR steering in a Bose-Hubbard system, it is preferable to start with a non-classical initial state and minimise collisional interactions, possibly using Feshbach resonance techniques~\cite{Feshbach}. In these systems, a quantum initial state plays a greater role than in nonlinear optical systems, where, for example, a squeezed pump was predicted to improve quantum correlations in travelling-wave second harmonic generation~\cite{Liz}. The essential difference is that second harmonic generation is perfectly capable of producing squeezed outputs even beginning from a completely classical pump. 

\section{Conclusions}
\label{sec:conclusions}

We have investigated the short time dynamics of the populations, entanglement and EPR steering in small Bose-Hubbard chains with a mesoscopic number of atoms initially inhabiting only the first well. The tunnelling is turned on at $t=0$, which is equivalent to a sudden change of parameters. Using correlations that are suitable for number conserving processes, we find that bipartite entanglement and EPR steering spreads along the chain and gives a periodic signal in the two- and three-well cases without interatomic interactions, but only when the initial state in the first well is sub-Poissonian. In the cases where we could obtain analytical solutions, we saw that the existence of the entanglement and EPR steering signals depended completely on the number statistics in the first well being sub-Poissonian. The Fock states we used are maximally sub-Poissonian, so that the signals were optimised for these cases. When we began with an initial coherent state, we saw no evidence of either entanglement or EPR steering in these cases. This is not unexpected since the linear tunnelling is somewhat analogous to a time dependent beamsplitter and it is known that this cannot produce entangled outputs from classical inputs. In the four-well system, the solutions for the populations and the different correlations were irregular and non-periodic, even without interactions. In this case an initial Fock state resulted in strong signals for entanglement and EPR steering between the two end wells.  

The solutions for the well populations did not change significantly over the time scales investigated, being virtually indistinguishable in the two- and three-well systems. The four-well population solutions did begin to diverge from the non-interacting populations after some time, but followed the same general form. The correlations for an initial Fock state however, were seriously affected by the interactions, only following the non-interacting solutions at short times. Some individual correlations which had signalled entanglement or EPR steering in the non-interacting case failed to do so once interactions were added. On the other hand, some correlations which failed to attain positive values for initial coherent states did become positive with interactions, although only over small time intervals and with small maximum values. In every case, the EPR steering correlations between any two wells were found to be different under a change of indices, with $\Sigma_{ij}$ and $\Sigma_{ji}$ never having positive values simultaneously. We cannot say if this would be the case for all possible EPR steering measurements, but any steering is asymmetric for the measure we used here. 

Overall, the initial quantum state in the first well was found to be far more important in achieving the bipartite correlations than the interactions. This suggests that if Bose-Hubbard dynamics are to be used for quantum information purposes, it would be preferable to lower the collisional interactions as far as possible and concentrate on producing non-classical initial states. As the natural state of the atoms in an isolated well is a Fock state, this may provide significant advantages.

\section*{Acknowledgments}

This research was supported by the Australian Research Council under the Future Fellowships Program (Grant ID: FT100100515). The author thanks Simon Haine for invaluable help with Mathematica.

\end{document}